\documentclass{article}
\usepackage{spconf,amsmath,graphicx}
\usepackage{hyperref, cite, amssymb, booktabs, relsize, multirow, subcaption}

\title{A DUAL-PATH FRAMEWORK WITH FREQUENCY-AND-TIME EXCITED NETWORK FOR ANOMALOUS SOUND DETECTION}
\name{Yucong Zhang$^{1,2}$ \qquad Juan Liu$^{1\dag}$\thanks{$^\dag$Corresponding Authors: Juan Liu: liujuan@whu.edu.cn, Ming Li: ming.li369@dukekunshan.edu.cn}  \qquad Yao Tian$^{3}$ \qquad Haifeng Liu$^{4}$ \qquad Ming Li$^{1,2\dag}$}
\address{
	$^1$School of Computer Science, Wuhan University, Wuhan, China\\
	$^2$Suzhou Municipal Key Laboratory of  Multimodal Intelligent Systems, \\ Duke Kunshan University, Kunshan, China \\
	$^3$Data \& AI Engineering System, OPPO, Beijing, China \\
	$^4$University of Science and Technology of China, Hefei, China \\
}
\begin{document}
%
\maketitle
\begin{abstract}
In contrast to human speech, machine-generated sounds of the same type often exhibit consistent frequency characteristics and discernible temporal periodicity. However, leveraging these dual attributes in anomaly detection remains relatively under-explored. In this paper, we propose an automated dual-path framework that learns prominent frequency and temporal patterns for diverse machine types. One pathway uses a novel Frequency-and-Time Excited Network~(FTE-Net) to learn the salient features across frequency and time axes of the spectrogram. It incorporates a Frequency-and-Time Chunkwise Encoder~(FTC-Encoder) and an excitation network. The other pathway uses a 1D convolutional network for utterance-level spectrum. Experimental results on the DCASE~2023 task~2 dataset show the state-of-the-art performance of our proposed method. Moreover, visualizations of the intermediate feature maps in the excitation network are provided to illustrate the effectiveness of our method.
\end{abstract}
\vspace{-4mm}\section{Introduction}\vspace{-3mm}
\label{sec:intro}
Anomalous sound detection~(ASD) is a task to distinguish anomalous sounds from normal ones. It is useful to monitor a machine's condition and detect malfunctions of an operating machine before it is damaged. ASD is a challenging task and is often regarded as an unsupervised learning problem~\cite{dcase2020}, given the rare occurrence and high diversity of anomalous events. Furthermore, in real-world scenarios, machines may operate under different settings and environmental conditions, leading to potential domain shifts~\cite{dcase2022, dcase2023, ToyADMOS2,MIMII_DG}, thereby increasing the difficulty of the ASD task.

To address the lack of anomalous data, conventional ASD systems adopt a generative method~\cite{Ellen2019Auto, Kaori2020IDNN} to model the distribution of normal data. Recently, self-supervised methods~\cite{Hadi2022Contrastive, Han2022Self, liu2022anomalous, zhang23j_interspeech} are getting more attention, which is widely adopted by top-ranked teams~\cite{LiuCQUPT2022, KuroyanagiNU-HDL2022, GuanHEU2022, JieIESEFPT2023, LvHUAKONG2023, JiangTHUEE2023, WilkinghoffFKIE2023} in recent DCASE\footnote{DCASE: Detection and Classification of Acoustic Scenes and Events, \url{https://dcase.community}} challenges. These systems train a feature extractor on normal data to obtain expressive embeddings, and use distance metrics to assess the abnormality by comparing test embeddings with normal ones. Despite the success of these systems, the frequency patterns and temporal periodicity remain relatively under-explored when modeling machine sounds.

Some recent studies have investigated the efficacy of frequency patterns in machine-generated sounds. In DCASE~2022 Challenge, the first-ranking team~\cite{LiuCQUPT2022} builds customized high-pass filters for individual machine types, enhancing ASD performance by applying them before the Mel filters. Additionally, experiments conducted by~\cite{MaiDGB22} demonstrate notable high-frequency characteristics produced by certain machine types. Nevertheless, these approaches rely on manually constructed filters to leverage frequency patterns, limiting their adaptability to new machine types.

To automatically explore the frequency patterns, one possible solution is to learn the patterns with deep learning. Recently, researchers in~\cite{zhang23fa_interspeech} have explored automated analysis of frequency patterns on top of their prior work~\cite{liu2022anomalous}. They introduce a multi-head self-attention~\cite{Vaswani2017AttentionIA} to adaptively filter the log-Mel spectrogram. Their experimental results demonstrate the feasibility of integrating frequency pattern analysis into the training process of ASD.

\begin{figure*}[t]
	\centering
	\includegraphics[width=1.01\textwidth]{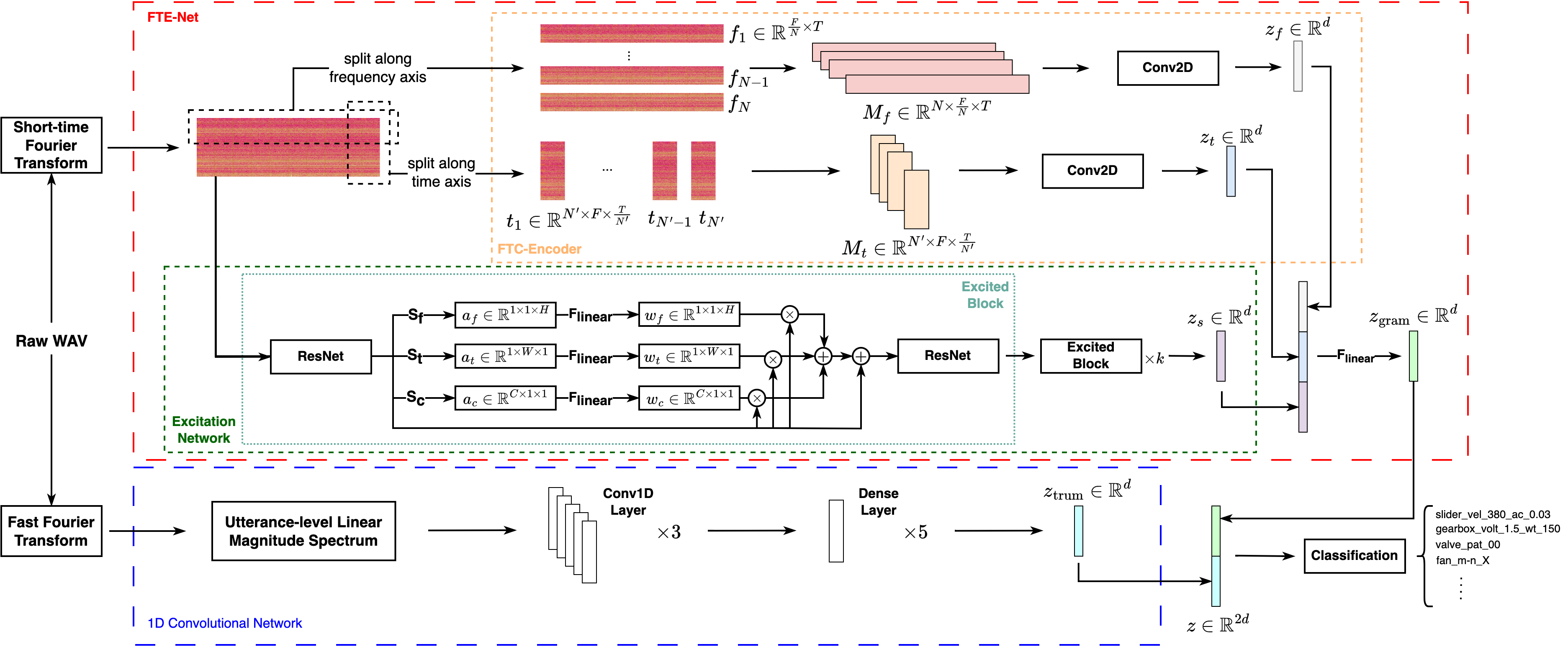}
	\vspace{-4mm}
	\caption{The overview of our proposed framework}
	\label{fig:overall_structure}\vspace{0mm}
\end{figure*}

In this paper, we propose a novel framework that leverages both the frequency and temporal characteristics. We use the framework from~\cite{Wilkinghoff2023DesignCF} as the backbone, dealing with both frame-level spectrogram and utterance-level spectrum. Different from~\cite{Wilkinghoff2023DesignCF}, we employ a Frequency-and-Time Excited Network~(FTE-Net) in the spectrogram pathway to enrich the learnt representation by capturing salient patterns in both the frequency and time domains. To the best of our knowledge, our work is the first to integrate both frequency and temporal pattern analysis of a spectrogram within a deep-learning framework for machine ASD.\vspace{0mm}

\vspace{-3mm}\section{Methods}\vspace{-3mm}
\label{sec:methods}

Our proposed framework uses~\cite{Wilkinghoff2023DesignCF} as the backbone, integrating a 1D convolutional network for utterance-level spectrum and an FTE-Net for frame-level spectrogram. The FTE-Net incorporates a Frequency-and-Time Chunkwise Encoder~(FTC-Encoder) and an excitation network. The overall structure of our method is depicted in Fig.~\ref{fig:overall_structure}. In section~\ref{subsec:backbone}, we introduce the backbone framework~\cite{Wilkinghoff2023DesignCF} and briefly explain the difference between theirs~\cite{Wilkinghoff2023DesignCF} and ours. In section~\ref{subsec:fte_net}, we introduce the proposed FTE-Net module and explain in detail the FTC-Encoder and the excitation network in the module.

\vspace{-4.5mm}\subsection{Backbone framework}\vspace{-2mm}
\label{subsec:backbone}
The backbone framework is a dual-path ASD framework~\cite{Wilkinghoff2023DesignCF}, designed to process both the frame-level spectrogram and utterance-level spectrum of machine-generated sounds in separate paths. The spectrum is processed by three 1D convolutional layers and five dense layers, and the spectrogram is processed by four ResNet~\cite{he2016deep} layers. Comparing to using only the spectrogram, empirical results from top-ranked teams~\cite{JieIESEFPT2023, WilkinghoffFKIE2023} show that by adopting such dual-path structure that handles the spectrogram and spectrum separately can produce better results.
In this work, we replace the network used in the spectrogram pathway of~\cite{Wilkinghoff2023DesignCF} with a novel FTE-Net, aiming to learn frequency and temporal patterns.

\vspace{-4.5mm}\subsection{Frequency-and-Time Excited Network~(FTE-Net)}\vspace{-2mm}
\label{subsec:fte_net}
The FTE-Net is a two-branch network. One branch employs an FTC-Encoder, and the other branch uses an excitation network. The FTC-Encoder allows the network to learn the potential patterns within small intervals of frequency or time, while the excitation network is used to filter out unrelated information and enhance the useful patterns in a global context.

\vspace{-3mm}\subsubsection{Frequency-and-Time Chunkwise Encoder~(FTC-Encoder)}\vspace{-1mm}
\label{subsubsec:ftchunk_encoder}

\begin{table}[h]
\vspace{-4mm}\centering
\caption{Structure of the Conv2D module in the FTC-Encoder. n indicates the number of layers or blocks, c is the number of output channels, k is the kernel size and s is the stride. h and w are the output height and width of the ResNet Blocks.}\vspace{-3mm}
\label{tab:freq_pathway}
\resizebox{.85\columnwidth}{!}{%
\begin{tabular}{@{}ccccc@{}}
\toprule
Operator     & n & c                   & k      & s      \\ \midrule
Conv2D 7x7   & 1 & 32                  & (7,7)  & (2,2)  \\
MaxPooling   & - & -                   & (3,3)  & (2,2)  \\
ResNet block & 4 & (64, 128, 128, 128) & (3,3)  & (2,2)  \\
MaxPooling   & - & -                   & (h, w) & (h, w) \\ \bottomrule
\end{tabular}%
}\vspace{-2.5mm}
\end{table}

The FTC-Encoder is designed to process spectrogram data in a chunkwise manner, with separate pathways dedicated to handle frequency chunks and time chunks respectively. The goal of this module is to capture potential patterns within short frequency bands and time intervals. 

In the frequency pathway, the input spectrogram $X\in \mathbb{R}^{F \times T}$ is equally segmented into $N$ overlapping frequency bands, denoted as $f_i \in \mathbb{R}^{\frac{F}{N} \times T}$. These frequency bands ${f_1, f_2, \cdots, f_N}$ are subsequently merged to create a band-wise 3D feature matrix $M_f \in \mathbb{R}^{N \times \frac{F}{N} \times T}$. Finally, $M_f$ is passed through a 2D convolution network~(as shown in Table~\ref{tab:freq_pathway}) to get the embedding $z_f\in \mathbb{R}^d$, with the number of chunks serving as the number of input channels. The first Conv2D and MaxPooling layer uses large kernel size, aiming to reduce the dimension of the input. The last MaxPooling layer is used to flatten the feature maps.

Similar strategies are applied to the dual pathway along the time axis, using the same structure after splitting the spectrogram into small time segments.

\vspace{-4mm}\subsubsection{Excitation network}
\label{subsubsec:excitation_net}\vspace{-2mm}

\begin{table}[t]
\centering
\caption{Structure of the excitation network with the same notations shown in Table.~\ref{tab:freq_pathway}.}\vspace{-3mm}
\label{tab:enhanced}
\resizebox{.75\columnwidth}{!}{%
\begin{tabular}{@{}ccccc@{}}
\toprule
Operator     & n                  & c                  & k      & s      \\ \midrule
Modified SE  & -                  & -                  & -      & -      \\
Conv2d       & 1                  & 16                 & (7,7)  & (2,2)  \\
MaxPooling   & -                  & -                  & (3,3)  & (2,2)  \\
Modified SE  & -                  & -                  & -      & -      \\ \midrule
ResNet Block & \multirow{3}{*}{1} & 16                 & (3,3)  & (1,1)  \\
Modified SE  &                    & -                  & -      & -      \\
ResNet Block &                    & 16                 & (3,3)  & (1,1)  \\ \midrule
ResNet Block & \multirow{3}{*}{4} & (32, 64, 128, 256) & (3,3)  & (2,2)  \\
Modified SE  &                    & -                  & -      & -      \\
ResNet Block &                    & (32, 64, 128, 256) & (3,3)  & (1,1)  \\ \midrule
MaxPooling   & -                  & -                  & (h, w) & (h, w) \\ \bottomrule
\end{tabular}%
}\vspace{-5mm}
\end{table}

The detailed structure of the excitation network is shown in Table~\ref{tab:enhanced}. Modified squeeze-and-excitation~(SE)~\cite{hu2018squeeze} modules are integrated between ResNet blocks to form the excited block. While the conventional SE generates a mask~($w_c$) to adjust channel-wise feature maps, we introduce two additional masks, namely the frequency excitation mask~($w_f$) and the time excitation mask ($w_t$). As shown in Fig.~\ref{fig:overall_structure}, given an input $x\in \mathbb{R}^{C\times H\times W}$, where $H$ and $W$ are the dimensions along the frequency and time axis, the excitation map is formulated as follows:
$$ w_i = \frac{1}{1+\exp\big(-(a_i\cdot W^{\operatorname{T}}+b)\big)},\quad a_i=S_{i}(x)
$$
where $S_{i}$ is a 2D average pooling operation, cancelling out the dimension other than $i$. $W$ and $b$ are learning parameters. The output is aggregated using the excitation maps as follows:
$$ y = x + \sum_{i\in \{c,f,t\}} w_i(x)\cdot x,
$$
where $w_c$, $w_f$, and $w_t$ represent the excitation masks for channel, frequency, and time respectively. 

As a result, the output embeddings~($z_f,z_t$) of the FTC-Encoder,  the embedding~($z_s$) of the excitation network are concatenated before passing to a linear layer to get the spectrogram embedding~($z_{\text{gram}}$). Meanwhile, the spectrum embedding~($z_{\text{trum}}$) is generated by the 1D convolutional network. To train the embeddings, $z_{\text{gram}}$ and $z_{\text{trum}}$ are stacked together, used as an input to a linear classifier to classify different machines.

\vspace{-3mm}\section{Experiments}\vspace{-1mm}
\label{sec:exp}

\vspace{-3mm}\subsection{Dataset}\vspace{-2mm}
\label{subsubsec:dataset}
The experiments are conducted on the DCASE 2023 Task 2 dataset~\cite{dcase2023}, which comprises audio clips from seven distinct machine types. Each machine type has roughly 1,000 audio clips, including 990 clips of source data and 10 clips of target data. Each audio clip lasts 6 to 18 seconds with a sampling rate of 16 kHz. The dataset includes a development dataset, an additional dataset, and an evaluation dataset. To compare with other systems in the challenge, 
the model is trained using the training portion of the development dataset and the additional dataset, while performance evaluation is conducted on the evaluation dataset. It is important to note that only normal machine sounds are used for training.
\vspace{-3mm}\subsection{Implementation details}\vspace{-2mm}
For data processing, we use linear magnitude spectrograms and spectrum as the inputs. The spectrogram is obtained by Short-time Fourier Transform, with the sampling window size and hop length set to 1024 and 512 respectively. The entire signal is used to obtain the utterance-level spectrum. In our experiments, we repeat and clip the audio to force its length to be 18 seconds.

In terms of 
the training strategy, we use the sub-cluster AdaCos~\cite{wilkinghoff2021sub} as the loss function to train the model. Wave-level mixup~\cite{zhang2018mixup} strategy is adopted as the data augmentation. We set the number of classes to match the joint categories of machine types and attributes. The model is optimized with the ADAM optimizer~\cite{adam} with a learning rate of 0.001. We set the batch size to 64 and train the model for 100 epochs.

The ASD results are generated by measuring the cosine distance between the prototypes of normal embeddings with the test embeddings for each machine type. Each machine type has 26 prototypes, including 16 center embeddings generated by K-Means on the source domain, and all the 10 embeddings from the target domain. 

The results are evaluated using the official scripts\footnote{Official scripts available at \url{https://github.com/nttcslab/dcase2023_task2_evaluator}}.  
Three commonly used metrics are adopted for evaluating the ASD performance in this paper: AUC, pAUC and the integrated scores. AUC is divided into source AUC and target AUC for the data in separate domains. pAUC is calculated as the AUC over a low false-positive-rate (FPR) range [0, 0.1]. The integrated score is the harmonic mean of AUC and pAUC across all machine types, which is the official score used for ranking.


\begin{table}[t]
\centering\vspace{-1mm}
\caption{Results (\%) on DCASE 2023 task 2 evaluation dataset. source AUC, target AUC, mean AUC and pAUC is the harmonic mean over all machine types. Integrated score is calculated using the official script$^2$.}\vspace{-2mm}
\label{tab:final_result}
\resizebox{.99\columnwidth}{!}{%
\begin{tabular}{@{}cccccc@{}}
\toprule
System &
  \begin{tabular}[c]{@{}c@{}}source\\ AUC\end{tabular} &
  \begin{tabular}[c]{@{}c@{}}target\\ AUC\end{tabular} &
  \begin{tabular}[c]{@{}c@{}}mean\\ AUC\end{tabular} &
  pAUC &
  \begin{tabular}[c]{@{}c@{}}Integrated\\ Score\end{tabular} \\ \midrule
Official baseline~\cite{Harada2023FirstshotAS} & -     & -     & 63.41 & 56.82 & 61.05 \\
Jie et al.~\cite{JieIESEFPT2023}        & -     & -     & 69.75 & 62.03 & 66.97 \\
Lv et al.~\cite{LvHUAKONG2023}         & -     & -     & 70.04 & 60.01 & 66.39 \\
Jiang et al.~\cite{JiangTHUEE2023}      & -     & -     & 68.03 & 60.71 & 65.40 \\
Wilkinghoff~\cite{WilkinghoffFKIE2023}       & -     & -     & 67.95 & 59.58 & 64.91 \\ \midrule
Self-implement Baseline        & $\mathbf{76.31}$ & 66.72 & 72.34 & 62.91 & 68.20 \\
Proposed Method using FTE-Net  & 72.94 & $\mathbf{75.08}$ & $\mathbf{73.97}$ & $\mathbf{66.38}$ & $\mathbf{71.27}$ \\ \bottomrule
\end{tabular}%
}\vspace{-6mm}
\end{table}

\vspace{-3mm}\subsection{Performance comparison and ablation studies}\vspace{-1mm}
We compare the performance of our proposed framework with the top~4 teams~\cite{JieIESEFPT2023, LvHUAKONG2023,JiangTHUEE2023,WilkinghoffFKIE2023} in the DCASE~2023 challenge. As presented in Table~\ref{tab:final_result}, our method surpasses all teams across all evaluation metrics. Notably, our approach exhibits a superior performance with a 4.3\% absolute improvement over the first-ranking team~\cite{JieIESEFPT2023} and a substantial 10.22\% absolute improvement over the official system~\cite{Harada2023FirstshotAS} in terms of the integrated score. The self-implement baseline shown in the table is re-implementation of~\cite{WilkinghoffFKIE2023} with more ResNet blocks added to the spectrogram branch. The results indicate that the proposed FTE-Net leads to improvements in the overall ASD performance.

Moreover, we find that the proposed framework exhibits a noteworthy capacity for domain generalization. As observed in Table~\ref{tab:final_result}, despite a moderate reduction in the source AUC compared to the baseline system, our framework demonstrates a substantial improvement in terms of the target AUC. We argue that the inferior performance of the source AUC is likely attributed to overfitting of the baseline system to the source data, given that the source and target domains feature a highly imbalanced ratio. An indicator of the overfitting phenomenon in the baseline system is the significant disparity between the source and target AUC values presented in the table. In contrast, the proposed FTE-Net exhibits a relatively minor difference, showing its generalization ability.

\begin{table}[h]
\centering\vspace{-2mm}
\caption{Results (\%) for different modules in FTE-Net.}\vspace{-2mm}
\label{tab:module_ablation}
\resizebox{.85\columnwidth}{!}{%
\begin{tabular}{@{}cccc@{}}
\toprule
System & \begin{tabular}[c]{@{}c@{}}mean\\ AUC\end{tabular} & pAUC & \begin{tabular}[c]{@{}c@{}}Integrated\\ Score\end{tabular} \\ \midrule
FTE-Net                                                                   & $\mathbf{73.97}$ & $\mathbf{66.38}$ & $\mathbf{71.27}$ \\ 
w/o FTC-Encoder              & 70.46 & 65.08 & 68.58 \\
w/o Excitation Network             & 71.78 & 64.18 & 69.06 \\ 
w/o Both~(Self-implement Baseline) & 72.34 & 62.91 & 68.20 \\ \bottomrule
\end{tabular}%
}\vspace{-5mm}
\end{table}

\begin{table}[h]
\centering\vspace{-2mm}
\caption{Results (\%) using different excitation mechanism.}\vspace{-2mm}
\label{tab:SE_ablation}
\resizebox{.65\columnwidth}{!}{%
\begin{tabular}{@{}cccc@{}}
\toprule
System & \begin{tabular}[c]{@{}c@{}}mean\\ AUC\end{tabular} & pAUC & \begin{tabular}[c]{@{}c@{}}Integrated\\ Score\end{tabular} \\ \midrule
FTE-Net & $\mathbf{73.97}$ & $\mathbf{66.38}$ & $\mathbf{71.27}$ \\
w/o Both~(Vanilla SE) & 69.94 & 62.31 & 67.20 \\
w/o freq. excitation & 69.43 & 64.48 & 67.79 \\
w/o time excitation  & 70.45 & 63.84 & 68.10 \\ \bottomrule
\end{tabular}%
}\vspace{-2mm}
\end{table}

To show the effectiveness of the individual modules in FTE-Net, we conduct some ablation studies. In Table~\ref{tab:module_ablation}, we show that the best performance is achieved by using all the modules. In Table~\ref{tab:SE_ablation}, we conduct an excitation mechanism ablation study. Our findings demonstrate that employing more excitation maps results in improved performance. Notably, frequency excitation maps outperform time excitation maps in terms of ASD performance.

\begin{figure}[t]
	\centering\vspace{0mm}
	\begin{minipage}[b]{.45\textwidth}
	\includegraphics[width=\textwidth]{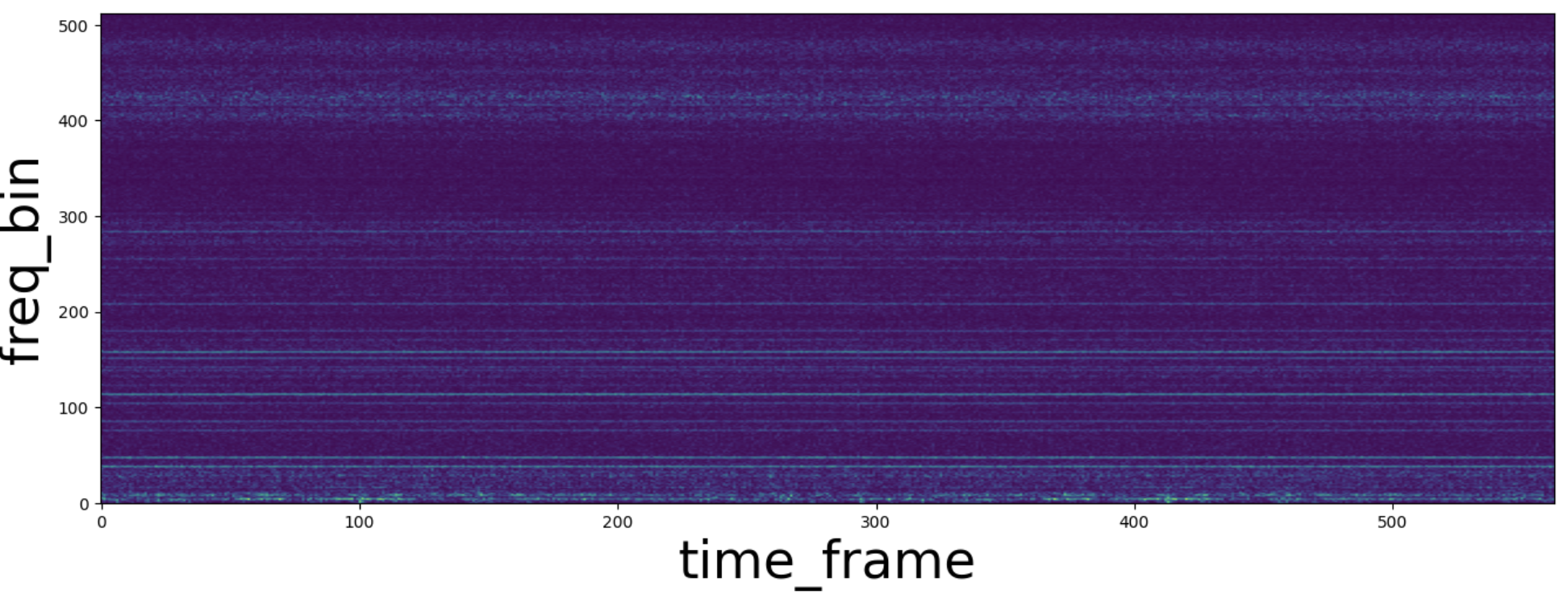}
	\subcaption{Origin spectrogram}
	\end{minipage}
	\vfill
	\begin{minipage}[b]{.45\textwidth}
	\includegraphics[width=\textwidth]{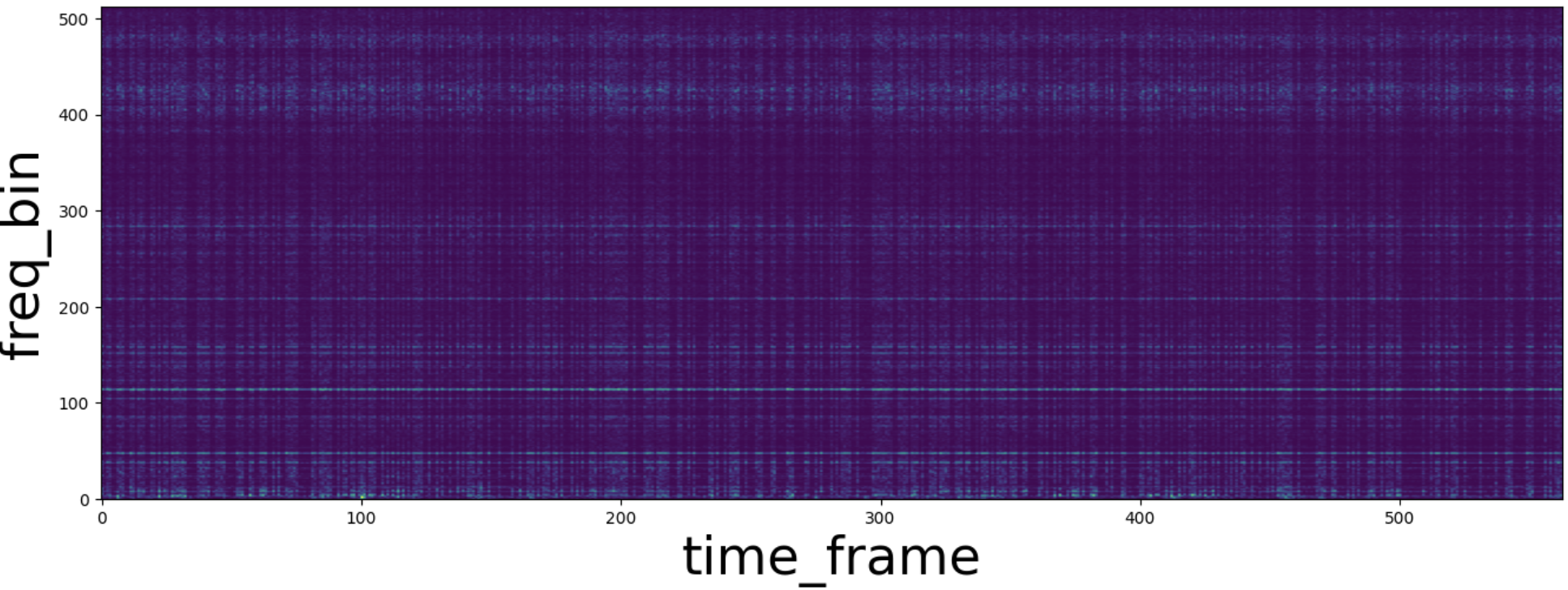}
	\subcaption{Feature map after the first excitation}
	\end{minipage}\vspace{-2mm}
	\caption{Illustration of the excited mechanism on fan}
	\label{fig:spec}\vspace{0mm}
\end{figure}

\begin{figure}[t]
	\centering
	\begin{minipage}[b]{.23\textwidth}
	\includegraphics[width=\textwidth]{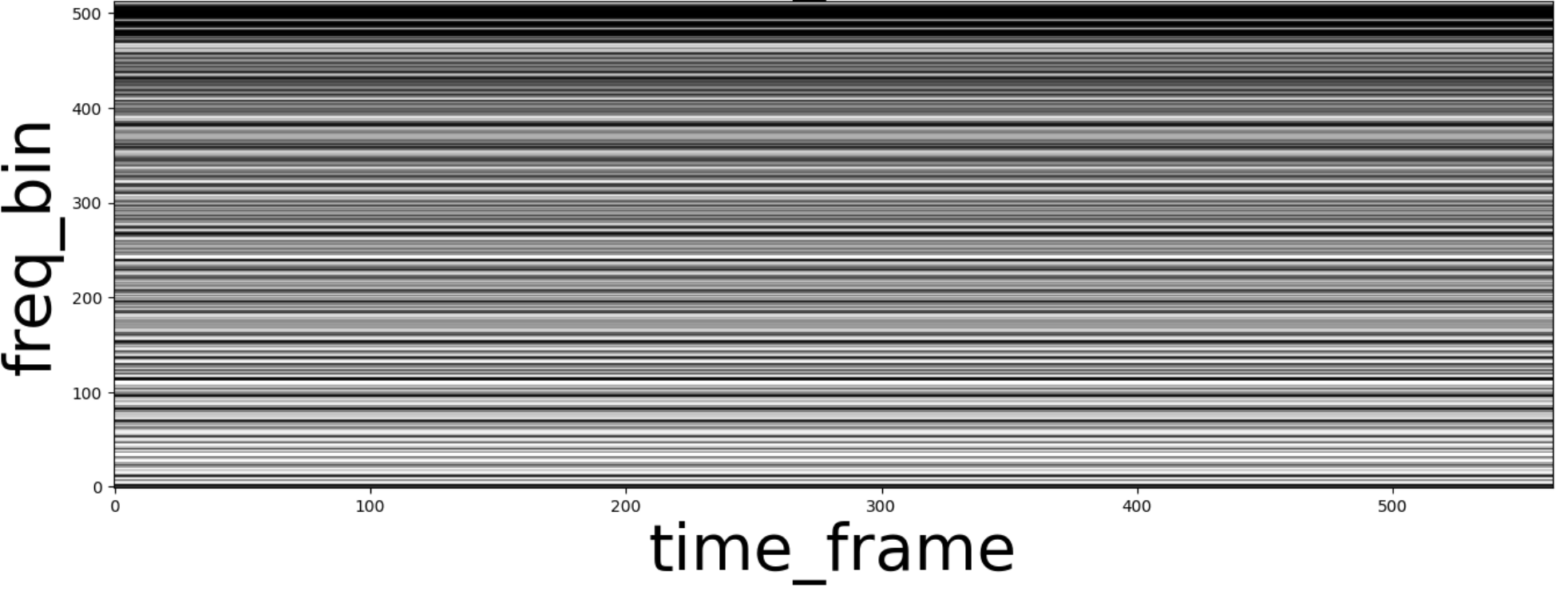}
	\subcaption{frequency-wise}
	\end{minipage}
	\hfil
	\begin{minipage}[b]{.23\textwidth}
	\includegraphics[width=\textwidth]{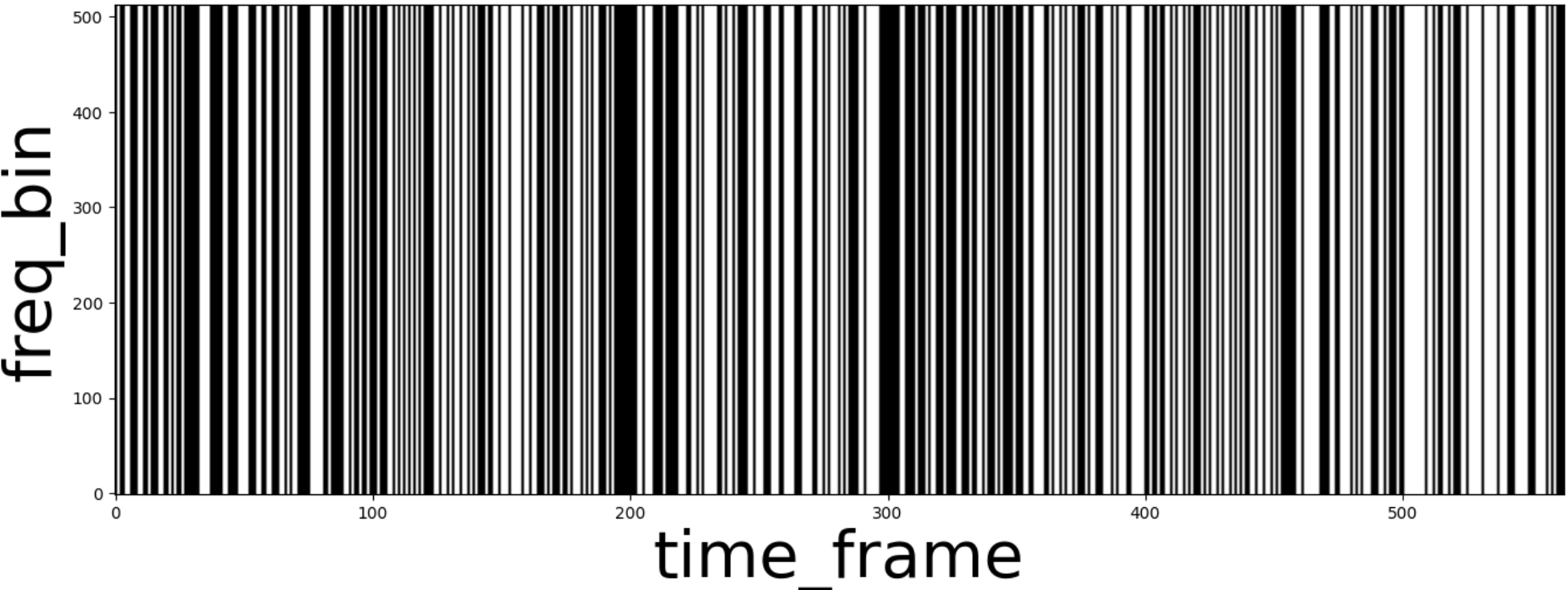}
	\subcaption{time-wise}
	\end{minipage}\vspace{-2mm}
	\caption{Illustration of excitation masks on fan}
	\label{fig:map}\vspace{0mm}
\end{figure}

\vspace{-3mm}\subsection{Visualization analysis}\vspace{-1mm}
To illustrate the impact of excitation mechanism in the excitation network, we present spectrogram comparisons before and after applying the excitation maps. In this example featuring a fan shown in Fig.~\ref{fig:spec}, we observe that the original spectrogram undergoes enhancement both in terms of frequency and time, highlighting the effectiveness of our method. Particularly, in the frequency excitation map shown in Fig.~\ref{fig:map}~(a), our network predominantly focuses on the high-frequency band, in accordance with the results given by recent discoveries~\cite{zhang23fa_interspeech, LiuCQUPT2022, MaiDGB22}. This indicates that our method effectively generates excitation maps conducive to machine sound modeling.

From Fig.~\ref{fig:map}~(a) and (b), frequency and temporal patterns can be highlighted. For example, despite the enhancement of the high frequency, some prominent frequency patterns in the middle range of the spectrogram are highlighted while some of them are filtered out. Additionally, despite the simple temporal periodicity, much more complicated temporal patterns within tiny time segments are shown. These patterns hold potential as features for analyzing sounds emitted by specific machine types in future research.

\vspace{-3mm}\section{Conclusion}\vspace{-3mm}
\label{sec:conclusion}
In our paper, we introduce a novel dual-path framework for anomaly detection in machine-generated sounds, which has the ability to leverage distinctive frequency and temporal patterns found in machine sounds. One pathway employs the Frequency-and-Time Excited Network (FTE-Net) to capture features across both frequency and time axes of the spectrogram. The other pathway utilizes a 1D convolutional network for utterance-level spectrum. The experiments on the DCASE~2023 task~2 dataset shows that our framework achieves state-of-the-art performance, demonstrating the effectiveness of leveraging dual attributes for machine ASD.


\vfill\pagebreak

\small{
\bibliographystyle{IEEEtran}
\bibliography{strings,refs}
}
\end{document}